\documentclass[graybox, envcountchap]{svmult}

\usepackage{mathptmx}        
\usepackage{amsmath}
\usepackage{ulem}
\usepackage{amssymb}
\usepackage{color}
\usepackage{helvet}          
\usepackage{courier}         
\usepackage{dirtree}
\usepackage{mathtools}

\usepackage{makeidx}        
\usepackage{graphicx}        
\usepackage{subfig}

\usepackage{multicol}        
\usepackage[bottom]{footmisc}

\usepackage{hyperref}        
\hypersetup{colorlinks=true,urlcolor=blue}

\usepackage[misc]{ifsym}

\makeindex             

\begin{document}


\title{Testing Gravity with Binary Black Hole Gravitational Waves}

\author{M.~Colleoni, N.V.~Krishnendu, P.~Mourier, S.~Bera, X.~Jiménez-Forteza}
\institute{Marta Colleoni (\Letter) \at Departament de F\'isica, Universitat de les Illes Balears, IAC3 -- IEEC, Crta. Valldemossa km 7.5, E-07122 Palma, Spain, \email{marta.colleoni@uib.eu}
\and N.~V.~Krishnendu \at International Centre for Theoretical Sciences (ICTS), Survey No. 151, Shivakote, Hesaraghatta, Uttarahalli, Bengaluru, 560089, India \email{krishnendu.nv@icts.res.in}
\and Pierre Mourier \at Departament de F\'isica, Universitat de les Illes Balears, IAC3 -- IEEC, Crta. Valldemossa km 7.5, E-07122 Palma, Spain, \email{pierre.mourier@uib.es}
\and Sayantani Bera \at Departament de F\'isica, Universitat de les Illes Balears, IAC3 -- IEEC, Crta. Valldemossa km 7.5, E-07122 Palma, Spain, \email{sayantani.bera@uib.eu}
\and Xisco Jiménez Forteza \at Departament de F\'isica, Universitat de les Illes Balears, IAC3 -- IEEC, Crta. Valldemossa km 7.5, E-07122 Palma, Spain, \email{f.jimenez@uib.es}
}

%
%
\maketitle

\abstract{
General Relativity (GR) remains the most accurate theory of
gravity to date. It has passed many experimental tests in the Solar
System as well as binary pulsar, cosmological 
and gravitational-wave (GW) observations. Some of
these tests probe regimes where gravitational fields are weak, the
spacetime curvature is small, and the characteristic velocities are
not comparable to the speed of light. Observations of compact
binary coalescences enable us to test GR in extreme environments of strong and dynamical gravitational fields, large spacetime curvature,
and velocities comparable to the speed of light. Since the breakthrough observation of the first GW signal produced by the merger of two black holes, GW150914, in September 2015, the number of confirmed detections of binary mergers has rapidly increased to nearly 100. The analysis of these events has already placed significant constraints on possible deviations from GR and on the nature of the coalescing compact objects. In this chapter, we discuss a selection of tests of GR applicable to observations of GWs from compact binaries. In particular, we will cover consistency tests, which check for consistency between the different phases of the binary's evolution, tests of GW generation, polarization and propagation, and tests of the remnant's nature. We conclude with a brief overview of the challenges and prospects for present and future observatories.}


\section{Introduction}
\label{sec:Introduction}

Coalescences of compact object (black holes and neutron stars) binaries are among the most violent events in the Universe and as such, are ideal sources of gravitational waves (GWs). The signals emitted by these processes fall under the category of transient GWs and can last between milliseconds and hundreds of seconds. Though such events take place far away from Earth, gravitational-wave interferometers are sensitive enough to detect their signatures, with the current detectors LIGO~\cite{LIGOScientific:2014pky}, Virgo~\cite{VIRGO:2014yos} and KAGRA~\cite{KAGRA:2020tym} (LVK) probing distances up to a few Gpc. Due to the steady improvement of these instruments~\cite{aLIGO:2020wna}, the detection rate of GWs emitted by compact binaries has been regularly increasing, with the LVK Collaboration announcing a total of 90 confirmed events as a result of the first three observing runs in 2015--2020~\cite{LIGOScientific:GWTC1,LIGOScientific:GWTC2,LIGOScientific:GWTC21,LIGOScientific:GWTC3}, along with a handful of additional detections announced by independent analyses (e.g., \cite{Nitz:2021,Olsen:2022,Nitz:2023,Mehta:2023}). This number is destined to grow further: the fourth observing run of the network started on 24 May 2023 (with several dozens of candidate events already announced~\cite{GraceDB}, and a fifth one (O5) is scheduled for the late 2020s~\cite{KAGRA:2013rdx,LIGO-T2000012-v1}. In preparation for each new observing run, GW detectors undergo systematic improvements, with the purpose of enhancing their sensitivity, range and stability, while minimizing the impact of transient noise artefacts, known as glitches, which can deteriorate data quality around transients of astrophysical origin~\cite{LIGOScientific:2019hgc}. Proposed upgrades of Advanced LIGO for the post-O5 era include the A$\sharp$ and Voyager~\cite{LIGO:2020xsf} configurations. The network of ground-based detectors is also set to expand: LIGO-India is expected to become operational in the first half of the 2030s \cite{LIGO-M1100296,Unnikrishnan:2013qw,Saleem:2021iwi}, and the third-generation ground-based detectors --- the Einstein Telescope \cite{ETsciencecase,Maggiore:2019uih} and Cosmic Explorer \cite{Reitze:2019iox} --- should become operational in the mid 2030s. Furthermore, several space-based GW observatories are also being planned for the next decade: LISA~\cite{LISA:2017pwj,Babak:2021mhe}, DECIGO~\cite{Kawamura:2020pcg,Seto:2001qf}, TianGO~\cite{Kuns:2019upi}, TianQin~\cite{TianQin:2015yph} and Taiji~\cite{Hu:2017mde}. 

Due the strong and highly dynamical gravitational field they generate, coalescences of compact binaries are prime candidates to carry out precision tests of gravity. Progress in analytical weak-field and gravitational self-force calculations, Numerical Relativity (NR) simulations, the effective-one-body formalism and black-hole (BH) perturbation theory (see~\cite{LISAConsortiumWaveformWorkingGroup:2023arg} for a recent review covering these frameworks) have propelled the development of ever more sophisticated models for the gravitational waveforms, tracking the dynamics of compact binaries and their GW signatures through the various stages of the coalescence, from the pre-merger (inspiral) to the post-merger (or ringdown) stage. The construction of such inspiral-merger-ringdown (IMR) models relies on the assumptions that GR is the correct theory of gravity and that environmental effects can be neglected (though these could be included \textit{a posteriori}, in a phenomenological way~\cite{CanevaSantoro:2023aol}). Hence, by definition, these templates will not be able to capture beyond-GR phenomena, nor the behaviour of exotic compact objects carrying extra fields, such as boson stars. It is also important to consider that specific GR templates may also rely on further assumptions regarding the nature of the coalescence which might not be accurate for all observed systems: many state-of-the-art gravitational waveform models, for instance, are restricted to quasi-circular inspirals~\cite{Pratten:2020ceb,Ramos-Buades:2023ehm,Varma:2019csw}. Though most binaries are expected to have almost fully circularised when they enter the detection band of current ground-based detectors, some systems might be characterised by a non-negligible eccentricity. GW templates for quasi-circular inspirals might yield artificial deviations from GR when applied to these systems \cite{Bhat:2022amc,Narayan:2023vhm,Saini:2023rto}. The development of GW models accounting for eccentricity is an active area of research (e.g., \cite{Nagar:2021gss,Nagar:2022fep,Ramos-Buades:2021adz}). 

Further modifications of the detected GW signals could be related to lensing effects. GWs can be lensed by clustered matter or isolated objects between the source and the detectors. Lensing could produce multiple images of the same signal in the detector, which will be time and phase-shifted with respect to each other and will have different amplitudes. Strong lensing is caused by massive lenses, such as galaxies and galaxy clusters~\cite{Oguri:2018muv}. It can lead to frequency-dependent time-delays between different harmonics of the GW strain which might have effects qualitatively similar to those induced by a modified dispersion relation~\cite{Ezquiaga:2020gdt,Ezquiaga:2022nak} (see Sec.~\ref{sec:propagation} below). Isolated objects acting as point-mass lenses might also distort GW signals (\textit{microlensing}~\cite{Takahashi:2003ix}), with multiple images interfering with each other. Lensing might open new avenues to carry out tests of GR with GWs~\cite{Baker:2016reh, Balaudo:2022znx, Chung:2021rcu, Collett:2016dey, Ezquiaga:2020dao,Goyal:2020bkm,Goyal:2023uvm}. We will not discuss further the possible interplay between lensing and beyond-GR effects and will assume that the GW signals detected so far are unlensed. This is consistent with the lack of evidence of lensing signatures in the data gathered so far~\cite{LIGOScientific:2021izm,LIGOScientific:2023bwz}.

While complete models for GWs emitted in alternative theories of gravity are now also under development, building upon progress on the post-Newtonian (see~\cite{Bernard:2022noq,Shiralilou:2021mfl,Trestini:2024zpi} for some recent works) and NR fronts (see, e.g.,~\cite{Foucart:2022iwu,LISAConsortiumWaveformWorkingGroup:2023arg,LISA:2022kgy,Okounkova:2019dfo,Ripley:2022cdh} and references therein), they have not reached the same level of maturity as their GR counterparts. Even without self-consistent templates accounting for these and other yet to be modelled physical effects, it is nevertheless still possible to carry out a number of tests of strong-field gravity with the purpose of detecting possible deviations from GR. These tests usually go under the name of null tests, where the null hypothesis being tested is that GR provides the correct description of gravity. 

One class of tests arising from such premises is that of consistency tests, which can be broadly divided into two categories: the first comprises residuals tests, which check whether the data are consistent with the GR waveforms being used, within noise fluctuations; the second includes internal consistency tests, whose goal is to assess whether different portions of the signal are consistent with each other (and, ultimately, with GR's predictions). 

Another strategy relies on adding parametric deviations to GR GW models or, more generally, to phenomenological models inspired by analytical results coming from weak-field techniques or black-hole perturbation theory. On a theoretical level, these parametric deviations could mimic a wide spectrum of phenomena not captured by the original, GR waveform models, e.g. a modified binary dynamics, the presence of scalar hairs, different horizon properties or modified dispersion relations distorting the signal's phase with respect to GR. Based on the physical content that is being probed, parametric deviations are implemented in different manners and over different frequency ranges. 

Finally, while GR only allows two tensorial polarization modes for GWs, the \textit{plus} $h_{+}$ and the \textit{cross} $h_{\times}$ mode, alternative theories of gravity might predict more. For instance, additional scalar polarization modes can appear in scalar-tensor theories; both vector and scalar modes are allowed in tensor-vector-scalar theories~\cite{Bekenstein:2005nv} and massive gravity~\cite{deRham:2014zqa}. Therefore, violations of GR could also be detected if there is evidence in the data of non-tensorial polarizations. 

In order to place stronger bounds on possible deviations from GR, single-event constraints are typically combined together. In a first approximation, one can assume that the value of such deviations is the same across all events. Under this assumption, combined probability distributions can be computed simply multiplying the likelihoods of several events. However, this might be a fairly strong requirement for those tests where it is reasonable to expect a certain dependence of the deviation from the source parameters. In order to relax the previous assumption, one can make the hypothesis that the observed deviations belong to a Gaussian distribution (with unknown mean $\mu$ and standard deviation $\sigma$), and can therefore take different values for different events. The posterior distribution for a specific deviation parameter, given a set of events, is then computed by marginalizing its posterior distribution over $\mu$ and $\sigma$. This approach is typically referred to as hierarchical combination~\cite{Zimmerman:2019wzo,Isi:2019asy}. When presenting combined bounds on deviations from GR, we will always refer to one of these two methods. 

In what follows, we will present a selection of tests of GR applicable to observations of GW from compact binaries. We will discuss consistency tests (Sec.~\ref{sec:consistency}), tests of GW generation (Sec.~\ref{sec:genTests}), polarization (Sec.~\ref{sec:polarization}) and propagation (Sec.~\ref{sec:propagation}) and tests of the remnant's properties (Sec.~\ref{sec:remnant}). This latter group encompasses tests of the GR no-hair theorem relying on the quasi-normal-mode (QNM) spectrum of the final remnant, as well as searches for ``echoes'' --- i.e. pulses of GW radiation which would be emitted during post-merger due to the lack of a perfectly absorbing event horizon. We refer the reader to Chapter 9 \cite{Franchini:2023eda} for an extensive treatment of BH QNMs and tests of GR based on the nature of the remnant object. We will conclude with a brief discussion of the challenges and future perspectives for tests of GR based on GWs (Sec.~\ref{sec:challenges}).

\section{Consistency tests}
\label{sec:consistency}

\subsection{Residuals tests}
\label{subsec:residuals}
Given a GR model for the GW signal emitted by a compact binary coalescence, we can ask ourselves how well it approximates the observed data. If GR is correct, and the model contains all the physical content that is relevant to describe the signal, the expectation is that the residual power left in the data after subtracting a fiducial best-fit template should be consistent with instrumental noise. Since the parameters of the binary are not known exactly, the best-fit parameters used in residuals analyses are usually those corresponding to the maximum-likelihood template for the signal, which are obtained through Bayesian parameter estimation techniques applied to the data. Notice that the maximum likelihood may or may not represent the true parameters of the binary. If $h(f)$ is a model template attempting to describe the true GW strain signal present in the data $d(f)$, in frequency domain, the maximum likelihood is obtained by minimizing the difference between the two computed as: 
\begin{equation}
    \log \mathcal{L} = -\frac{1}{2}\sum_{i} \, \langle d_i-h_i \rangle^2 \; ,
    \label{eqn:logl}
\end{equation}
for the $i$-th detector recording the data $d_i$, and the corresponding strain template $h_i$. One important quantity to note here is the noise-weighted inner product, which in frequency domain reads
\begin{equation}
    \langle a,b \rangle =4 \, \mathcal{R}\left[ \int_{f_\mathrm{low}}^{f_\mathrm{high}} \, \frac{a(f) \, b(f)^{\ast}}{S_n (f)}\, df \right] \; ,
    \label{eqn:innerproduct}
\end{equation}
where $\mathcal{R}$ denotes the real part, the ${}^{\ast}$ sign the complex conjugate of the frequency-domain functions and $S_n (f)$ is the power spectral density of the detector. The integral's lower and upper cut-off frequencies $f_\mathrm{low}, f_\mathrm{high}$ are determined by the detector's sensitivity and the signal's duration.
In the ideal case where the template $h(f)$ exactly matches the data $d(f)$, and no noise is present, the inner product in Eq.~\eqref{eqn:logl} vanishes. However, in reality, $d(f)$ will include both a signal and a noise component and the template $h(f)$ will not exactly match the signal: therefore the inner product in Eq.~\eqref{eqn:logl} will not be zero, and there will be residual power left after the subtraction. The residual analysis looks for the presence of such coherent signal in $d_i-h_i$ --- which, if statistically significant, might indicate a deviation to the GR waveform model used --- employing 
algorithms such as \textsc{BayesWave}~\cite{Cornish:2014kda}, which can model coherent data features across detectors using sums of Morlet-Gabor sine-Gaussian wavelets, whose parameters are estimated through stochastic sampling.  %

The strength of the residual term is measured in terms of its signal-to-noise-ratio (SNR). Before defining the residual SNR, let us start with the optimal SNR, which is obtained by evaluating the inner product in Eq.~\eqref{eqn:innerproduct} by assuming $h_i$ and $d_i$ are the same and taking a 2-norm over the detectors. If we assume $h_i$ is the best-fit GR template $h_\mathrm{GR}$, we get the optimal SNR for $h_\mathrm{GR}$, $\mathrm{SNR}_\mathrm{GR}$:
\begin{equation}
    \mathrm{SNR}_\mathrm{GR} = \left( \sum_i \langle h_\mathrm{GR,i}, h_\mathrm{GR,i} \rangle^2 \right)^{1/2}.
\end{equation}
Similarly, the residual SNR is calculated using the same formula but for the quantity $d_i-h_{\mathrm{GR},i}$. For each event, \textsc{BayesWave} returns a distribution of possible models for the residuals, each with its own SNR: this allows to establish a fiducial upper limit on the coherent power that is unaccounted for by the GR model. Conventionally, the upper limit is taken to be the 90\%-credible upper limit and is denoted as $\mathrm{SNR}_{90}$.
To check that the values of $\mathrm{SNR}_{90}$ are consistent with detector noise, the same analysis is then run on hundreds of noise-only stretches of data near the time of each detection, yielding a 90\%-credible upper limit on the SNR of coherent power with an instrumental origin, $\mathrm{SNR}_{90}^{n}$. Finally, a probabilistic statement is made, quantifying the probability that $\mathrm{SNR}_{90}^{n}$ is larger than $\mathrm{SNR}_{90}$ through a p--value   $p=P(\mathrm{SNR}_{90}^{n}\geq \mathrm{SNR}_{90}|\mathrm{noise})$, with lower p--values indicating a lower probability that the coherent power in the residuals is due to instrumental noise only. 

\begin{figure}[h]
\begin{center}
\includegraphics[width=1.\textwidth]{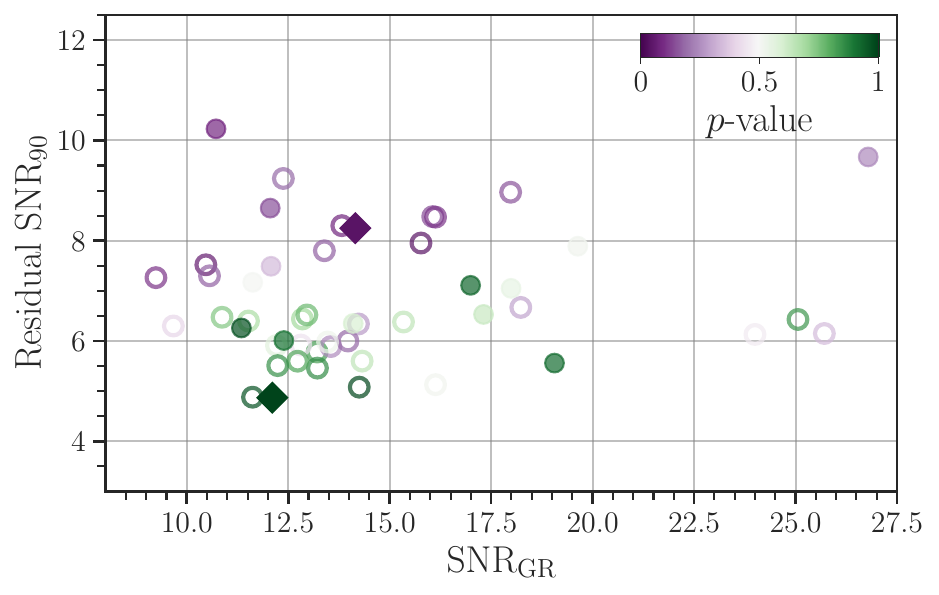}
\caption{The residual SNR and the optimal SNR from the best-fit template (assuming GR is the correct theory) are shown for the events from the third gravitational wave transient catalogue (GWTC-3)~\cite{LIGOScientific:GWTC3}, covering all events published by the LVK Collaboration so far. 
The $p$-values from individual events are indicated by the color scheme in the top-right corner. The highest $p$-value is marked with a green diamond, and the lowest $p$ value is marked with a purple diamond. Filled circles represent results from the second half of the third observation run (O3-b), whereas empty circles represent all events detected previously. As expected, if the GR waveform model used accurately describes the observed signals, no correlation is visible between the optimal SNR and the residual SNR of the events. Likewise, the expected ---uniform--- distribution of $p$--values can be visually inferred from the distribution of green and purple points. This figure is taken from~\cite{LIGOScientific:2021sio}.}
\label{fig:res-o3b}
\end{center}
\end{figure}

Assuming that GR is the correct theory of gravity (or, more precisely, that the GR-based templates and noise model employed in parameter estimation are sufficiently accurate for the purpose of the test~\cite{LIGOScientific:2019hgc}), we should not observe any clear correlation between $\mathrm{SNR}_{90}$ and $\mathrm{SNR}_{\mathrm{GR}}$; furthermore, p--values should be uniformly distributed between 0 and 1. These two trends are indeed observed in the data in the analyses performed by the LVK Collaboration so far~\cite{LIGOScientific:2019,LIGOScientific:2020tif,LIGOScientific:2021sio}, as illustrated in Fig.~\ref{fig:res-o3b}. The lack of a clear correlation between $\mathrm{SNR}_{\mathrm{GR}}$ and $\mathrm{SNR}_{90}$ is a sign that the power not accounted for by the GR model is likely due to noise fluctuations, as systematic deviations would result in higher $\mathrm{SNR}_{90}$ for increasing values of $\mathrm{SNR}_{\mathrm{GR}}$ (i.e., systematic inaccuracies in the GR models should become more apparent for louder signals).

\subsection{Inspiral-merger-ringdown consistency tests}
\label{subsec:imr_consistency}

The inspiral-merger-ringdown (IMR) consistency tests aim at establishing whether the remnant's mass and spin inferred from the inspiral part of the signal are consistent with those extracted from the post-inspiral part of the signal. This induces a test of GR's black-hole no-hair and uniqueness theorems, together with the astrophysical assumption of electrically neutral BHs, which imply that the final remnant should be a Kerr BH, fully characterized by its two parameters: mass and spin. The remnant's parameters can be extracted independently in the two regions, provided the signal accumulates enough SNR in each of them~\cite{Ghosh:2016qgn}. The LVK Collaboration, for instance, sets a SNR threshold of 6 that the signal needs to accumulate separately in each region in order to be considered for the test~\cite{LIGOScientific:2019,LIGOScientific:2020tif,LIGOScientific:2021sio}. This preliminary selection can be informed by a fiducial GR-model parameter estimation run on the full signal\footnote{%
In this case, the fiducial values considered for the SNR calculation are typically the maximum \textit{a posteriori} values of a full IMR fiducial parameter estimation run.%
}. 

The final mass $M_{\rm{f}}$ and spin $\chi_{\rm{f}}$ of the remnant can be estimated from the individual parameters of the progenitor binary, with the aid of fits calibrated to NR simulations (e.g., \cite{Healy:2014,Zlochover:2015,Hofmann:2016,JimenezForteza:2017,Boschini:2023,deLlucPlanas:2024}). 
Deviations from GR can be constrained through the fractional differences
\begin{eqnarray}
\dfrac{\Delta M_{\mathrm{f}}}{\bar{M}_{\mathrm{f}}}&=2 \,\dfrac{M_{\mathrm{f}}^{\mathrm{insp}}-M_{\mathrm{f}}^{\mathrm{post-insp}}}{M_{\mathrm{f}}^{\mathrm{insp}}+M_{\mathrm{f}}^{\mathrm{post-insp}}},\\
\dfrac{\Delta \chi_{\mathrm{f}}}{\bar{\chi}_{\mathrm{f}}}&=2\,\dfrac{\chi_{\mathrm{f}}^{\mathrm{insp}}-\chi_{\mathrm{f}}^{\mathrm{post-insp}}}{\chi_{\mathrm{f}}^{\mathrm{insp}}+\chi_{\mathrm{f}}^{\mathrm{post-insp}}},
\label{eq:imr_deviations}
\end{eqnarray}
where $\bar{M}_{\mathrm{f}}$ and $\bar{\chi}_{\mathrm{f}}$ denote the mean of the inspiral and post-inspiral estimates for the remnant's parameters. Values of $\Delta M_{\mathrm{f}}/\bar{M}_{\mathrm{f}}$ and $\Delta \chi_{\mathrm{f}}/\bar{\chi}_{\mathrm{f}}$ consistent with 0 indicate agreement with the predictions of GR. 

In LVK analyses, this test is carried out in the frequency domain, following the method outlined in Ref.~\cite{Ghosh:2016qgn}, where the transition between the two regimes is taken to coincide with the frequency of the GW quadrupolar mode at the innermost stable circular orbit (ISCO), for a Kerr remnant BH with mass and spin estimated analysing the full signal. Furthermore, LVK analyses are restricted to events with a detector-frame total mass below a fiducial threshold, to ensure enough SNR is accumulated in the inspiral.

\begin{figure}[h!]
\begin{center}
\includegraphics[width=\textwidth]{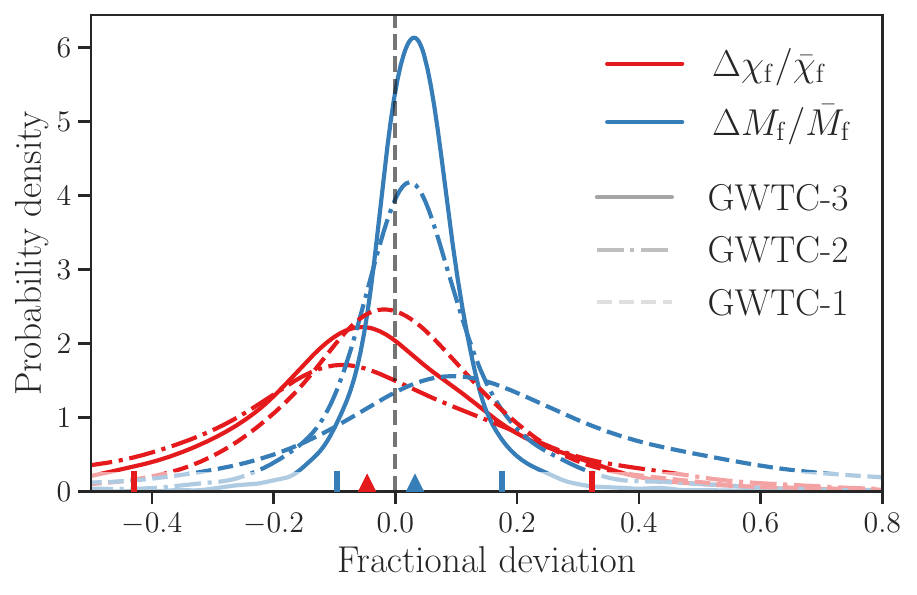}
\caption{Posterior distributions for the fractional differences of Eq.~\eqref{eq:imr_deviations}, quantifying possible discrepancies between the inspiral and post-inspiral measurements of the remnant's mass (blue curves) and spin (red curves). Different line styles indicate results hierarchically combining events from different observing runs of the LIGO-Virgo network. Triangles and bars on the x-axis denote the medians and 90\%-credible intervals of the GWTC-3 results. A value of 0 indicates perfect agreement between inspiral and post-inspiral estimates. This figure is taken from~\cite{LIGOScientific:2021sio}.\label{fig:imrct_hier}}
\end{center}
\end{figure}

Analyses performed so far have not shown any statistically significant disagreement between the inspiral and post-inspiral measurements. Fig.~\ref{fig:imrct_hier} shows the combined posterior distributions for the fractional deviation parameters $\Delta M_{\mathrm{f}}/\bar{M}_{\mathrm{f}}$ and $\Delta \chi_{\mathrm{f}}/\bar{\chi}_{\mathrm{f}}$ obtained by the LVK Collaboration on GWTC-3 data (corresponding to the three observing runs O1-O2-O3). Red and blue curves correspond to posteriors for the combined fractional deviations in the remnant's spin and mass, respectively. In both cases, zero is included in the 90\%-credible intervals (marked by solid thick lines on the x-axis), indicating good consistency between the two estimates, and therefore no clear deviations from GR. 

A drawback of this approach is the possibility of spectral leakage between the two regions, with power emitted during the inspiral leaking into the post-inspiral frequency band, and \textit{vice versa}. Ref.~\cite{Ghosh:2017gfp} estimated that spectral leakage is negligible for an aligned-spin system undergoing a quasi-circular inspiral and with the above choice of transition frequency. This follows from the observation that such signals have a monotonically increasing frequency and hence the application of the stationary phase approximation~\cite{Tichy:1999pv} (SPA) allows to identify the low-frequency and high-frequency parts of the signal with the inspiral and post-inspiral regimes respectively. However, a clean separation cannot be guaranteed for those systems exhibiting a non-monotonic frequency evolution, such as eccentric or precessing binaries \cite{Ashton:2022ztk}. To circumvent this problem, the signal can alternatively be split in the time domain~\cite{Ashton:2022ztk, Kastha:2021chr}. 

The choice of the exact time or frequency marking the transition between the regions is somewhat discretionary. Ref.~\cite{Kastha:2021chr} considered a number of alternatives, including the minimum-energy circular orbit and the merger time. The same reference explored the use of different (GR-based) waveform models for the inspiral and post-inspiral regimes to investigate the sensitivity of the test to the small differences between those models, concluding that they yield consistent results for the events considered.

\subsection{Test of the area theorem}
The second law of the BH thermodynamics states that the horizon area of a classical BH must increase over time~\cite{Hawking:1971tu}: this can be viewed as a fundamental consequence of the cosmic censorship hypothesis in GR~\cite{Penrose:1969pc,chruściel2000regularity}. In GR, the area of an astrophysical (uncharged) BH is,
\begin{equation}
    A(M_{\mathrm{BH}},\chi)=8 \pi \left(\frac{G M_{\mathrm{BH}}}{c^2}\right)^2 \left(1 + \sqrt{1-\chi^2}\right)\,
    \label{eq:bh_area}
\end{equation}
where $M_{\mathrm{BH}}$ and $\chi = J c/(G M_{\mathrm{BH}}^2) \in [-1,1]$ 
are the BH's mass and spin (i.e., the dimensionless version of its angular momentum $J$), respectively. The BH no-hair and uniqueness theorems in GR imply that the astrophysical BH properties depend only upon its mass $M$ and spin $\chi$. 
\begin{figure}[h!]
\begin{center}
\includegraphics[width=\textwidth]{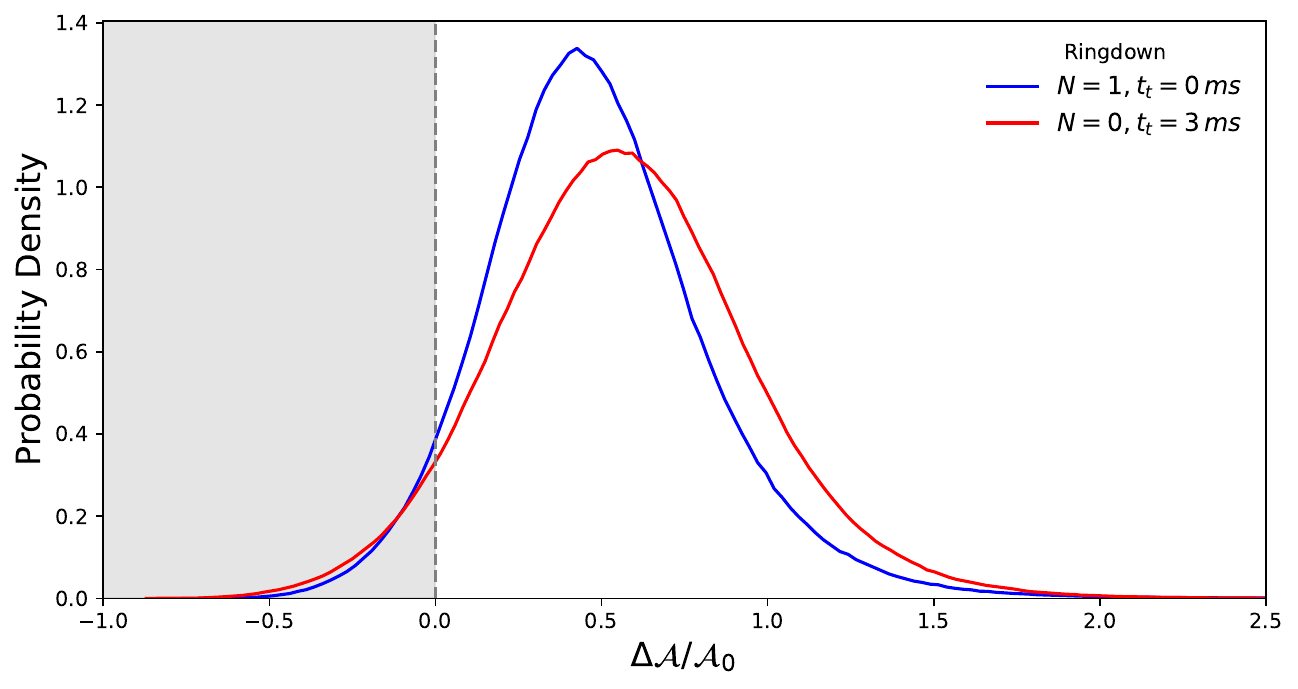}
\caption{Test of the BH area theorem using the posterior distributions presented in~\cite{Kastha:2021chr}. The blue curve provides the results obtained for $A_\mathrm{f}$ using a ringdown model made of one short-lived mode (\textit{overtone}; $N=1$) in addition to the dominant mode, starting at a truncation time $t_t = 0\, \text{ms}$ (with respect to the time of merger). The red curve shows the corresponding results using a dominant-mode-only ($N=0$) ringdown model starting at a truncation time $t_t = +3\, \text{ms}$. The gray area represents the region forbidden by the BH area theorem, which is excluded at the $\sim 94\%$ credibility level for the $N=1$, $t_t = 0 \, \text{ms}$ model.\label{fig:bh_area_th}}
\end{center}
\end{figure}
In a binary BH merger, one can recognise three different regimes, namely, the inspiral, merger, and ringdown regimes. In the inspiral-merger regime, and by virtue of the no-hair theorem, the area of both progenitor BHs is given in terms of their individual masses and spins and we denote their sum by $A_0 \equiv A(M_{1},\chi_{1})+A(M_{2},\chi_{2})$. In the ringdown regime, a final common horizon described solely by the final mass $M_\mathrm{f}$, the final spin $\chi_\mathrm{f}$ is formed with an area $A_\mathrm{f} \equiv A(M_\mathrm{f},\chi_\mathrm{f})$. Provided GR is the correct theory of gravity, the  BH area theorem states that
\begin{equation}
  A_\mathrm{f} > A_0 \,.
    \label{eq:bh_area_th}
\end{equation}
GWs provides us an excellent channel to test the area theorem. On the one hand, one can perform parameter inference on the inspiral-merger portion of the signal to obtain the marginalised posterior distributions for the progenitor parameters $M_{1,2},\chi_{1,2}$, which determine $A_0$. On the other hand, the same exercise can be repeated on the ringdown regime to infer the values of the final mass and final spin $M_\mathrm{f},\chi_\mathrm{f}$, which give $A_\mathrm{f}$. With that information in hand, a normalised version of the theorem can be expressed as,
\begin{equation}
    \frac{\Delta A}{A_0} > 0 \,,
    \label{eq:bh_area_th_norm}
\end{equation}
with $\Delta A = A_\mathrm{f} -A_0$. The accuracy and validity of this test relies strongly on the amount of SNR accumulated both from the inspiral-merger and the ringdown regime separately. However, as it may occur in the IMR consistency tests, the results may be subjected to systematic errors sourced by different choices of the truncation time $t_t$ that separate the two regimes, where $t_t=0$ is the time at which the peak of the strain is observed. The use of different templates has an impact on the probability at which the area theorem has been confirmed, with the exact result being sensitive to the IMR model employed before merger, and also to the number of damped-sinusoid modes (the so-called \textit{quasi-normal} modes) used to model the ringdown phase. The analysis of~\cite{Isi:2020tac} confirmed the area theorem with at least 95\% probability (or higher, depending on the number of ringdown modes added to the post-merger template).
The analysis of~\cite{Kastha:2021chr} showed a similar area theorem confirmation, at $\sim 94\%$ confidence or higher, although with an opposite trend in the evolution of this constraint with the inclusion of short-lived modes (\textit{overtones}) in the ringdown model.
Fig.~\ref{fig:bh_area_th} shows the posterior distributions for the fractional deviation $\Delta A/A_0$ for the analysis of the first GW event GW150914, using data from~\cite{Kastha:2021chr}. The figure illustrates the effect of different choices of the number of ringdown modes included in the post-merger template, as well as of the truncation time separating the pre-merger and post-merger regimes.

\section{Gravitational wave generation tests}
\label{sec:genTests}

\subsection{Parametrized generic modifications}
\label{subsec:generic_modifications}

In the absence of a preferred alternative theory of gravity, deviations from GR can be treated in a model-agnostic way, by introducing parametric deviations from GR in a fundamental object of the theory (such as the metric tensor) or directly in some observable quantities. Parametrized tests have a long history, with several formulations targeting different observational contexts: examples include the parametrized post-Newtonian formalism~\cite{Will:1972zz,Will:2014kxa}, which was designed to perform tests of GR in the weak-field regime, and the parametrized post-Keplerian formalism~\cite{Damour:1991rd}, conceived to test GR through binary pulsar observations (see also Ref.~\cite{Sampson:2013wia} for a comparison of different parametrized formalisms).  

Here, we will focus on parametrized tests specifically designed for interferometric GW detectors, where deviations are directly applied to the frequency-domain waveform models employed in GW data analysis~\cite{Arun:2006yw}, in a theory-agnostic way. This strategy has been translated into a number of frameworks that allow to place bounds on possible deviations from GR through GW observations, such as the parameterized post-Einsteinian framework~\cite{Yunes:2009ke}, the flexible theory-independent formalism~\cite{Mehta:2022pcn} and the Test Infrastructure for GEneral Relativity (TIGER) framework~\cite{Agathos:2013upa,Li:2011cg}. While sharing the same core principle, these frameworks differ in practical aspects of the implementation of the test, such as the number of simultaneous deviations allowed, the parameters that are being perturbed or the frequency range over which the deviations are activated.

Generically, one expects possible deviations from GR to be present both in the amplitude and phase of the signal. However, analyses often focus on modifications of the signal's phasing which can accumulate over time. This is especially relevant for binary systems inspiralling within the detector's frequency band over a sizeable number of cycles.

For this reason, we will first discuss the possibilities of performing parametrized tests in the weak-field regime, where the post-Newtonian (PN) formalism provides an accurate description of the dynamics of compact binaries. This applies when the two compact objects are still sufficiently far apart, i.e., early enough in the inspiral phase. The PN formalism is an analytic approximation scheme where the metric components and the associated gravitational fields are perturbatively expanded in powers of a velocity parameter $v/c$, with terms scaling like $(v/c)^{n}$ being referred to as of ``order $(n/2)$PN''. 

Here, we will consider aligned-spin systems in a quasi-circular inspiral. In this case, the SPA~\cite{Tichy:1999pv} yields closed-form expressions for the GW strain as a function of frequency:
\begin{equation}
    h(f)=A(f)\,e^{i\phi(f)},
\end{equation}
where $A(f)$ is the amplitude of the signal and $\phi(f)$ is its phase. Both amplitude and phase are expansions in the PN parameter starting from the leading (Newtonian) order. We will neglect amplitude corrections, given that, at current sensitivities, GW data analysis is mostly sensitive to phase variations, and we will focus on the signal's phase, which takes the following PN-expanded form:
\begin{equation}
    \phi(v) \sim \frac{3}{128\eta (v/c)^5} \sum_{n=0}^{N} \big(\phi_n+ \phi_{n\ell}\log(v/c) \big) (v/c)^{n}.
    \label{eq:phase_gr}
\end{equation}
In the above equation, $\eta=m_1 m_2/(m_1+m_2)^2$ denotes the symmetric mass ratio, where $m_1$ and $m_2$ are the component masses, and the PN expansion parameter $v/c$ can be re-expressed in terms of the total mass $M=m_1+m_2$ and the GW frequency $f$ as $v/c= c^{-1} (\pi G M f)^{1/3}$. 
The sum runs from the leading order $n=0$, corresponding to the Newtonian order, to the maximum PN order available to date, which is $3.5$PN when including spin effects\footnote{%
The phasing for non-spinning binaries, on the other hand, is known up to $4.5$PN order~\cite{Blanchet:2023bwj}; see also~\cite{Cho:2022syn} for a computation of spin effects on the phasing at $4$PN order via an effective field theory framework.%
}, with logarithmic terms appearing ($\phi_{n\ell} \neq 0$) for $n=5,6$, i.e., at $2.5$PN and $3$PN order, respectively. 

In GR, the PN coefficients $\phi_n$, $\phi_{n\ell}$ for binary BHs are entirely determined by the masses and spins of the binary (and, in the presence of neutron stars, by extra parameters describing their deformability properties)\footnote{In general, the PN coefficients might also depend on the binary's orbital eccentricity. However, we assume quasi-circular inspirals here.}. 
For instance, the leading-order coefficient reads $\phi_0=1$, the $0.5$PN coefficient $\phi_1$ is exactly 0, the $1$PN coefficient is a function of the component masses: $\phi_2=\left(3715/756+(55/9)\eta\right)$, and so on~\cite{Blanchet:2023bwj}. Following linear momentum conservation, in GR the dipole gravitational radiation vanishes. On the other hand, some alternative theories of gravity predict the presence of dipolar radiation. For example, in the Jordan-Fierz-Brans-Dicke (JFBD) scalar-tensor theory~\cite{Will:2018bme}, the phasing of Eq.~\eqref{eq:phase_gr} will feature a non-vanishing contribution at ($-1$)PN order (corresponding to $n=-2$), which, at leading order in the scalar-tensor coupling of the theory, $\alpha_0$, reads: 
\begin{equation}
    \phi_{-2} = -\frac{5(\alpha_1 - \alpha_2)^2}{168}+\mathcal{O}(\alpha_0^{4}),
\end{equation}
where $\alpha_i$ are the scalar charge  parameters of the body~\cite{Mehta:2022pcn}. More generally, we expect the PN coefficient structure to be different in alternative theories of gravity, compared to GR.  

Let us look into the practical implementation of parameterized tests of the GW phasing, starting from the inspiral region, where the gravitational wave phase takes the analytical form in Eq.~\eqref{eq:phase_gr}. First, one can introduce parametrized deviations in the individual phasing coefficients, e.g. 
\begin{equation}
\phi_{n}\xrightarrow{}\phi_{n}+\delta\phi_{n},
    \label{eq:ppn1}
\end{equation}
where $\delta\phi_{n}$ accounts for any possible deviation from GR at a specific PN order. Note that, though $\delta\phi_{n}$ could depend on the parameters describing the binary and the specific theory considered, its exact form is not needed for the purpose of the test.
Clearly, $\delta\phi_{n}=0$ if GR is the correct theory of gravity (or if no deviations are expected at a specific order). Since each PN coefficient will scale differently with the binary's parameters, for those terms in the series~\eqref{eq:phase_gr} that are non-zero, it is possible to introduce fractional deviations, rewriting Eq.~\eqref{eq:ppn1} as
\begin{equation}
\phi_{n}\xrightarrow{}\left(1+\frac{\delta\phi_{n}}{\phi_{n}}\right)\phi_{n}\equiv\left(1+\delta\hat{\phi}_{n}\right)\phi_{n}.
\label{eq:ppn2}
\end{equation}

Generic alternative theories of gravity will exhibit modifications at multiple PN orders, though these deviations might be highly degenerate. Hence, the constraints placed by multi-parameter tests, where multiple deviations are allowed at the same time, will be weaker than those obtained with single-parameter tests (where only one coefficient is varied, with the others being fixed at their GR value), and generally uninformative at current sensitivities~\cite{GW150914_TGR}. Furthermore, the presence of beyond GR effects would still be detectable when varying individual coefficients, even though the mapping to specific theories would not be possible in this case. For this reason, recent analyses performed by the LVK Collaboration~\cite{LIGOScientific:2019,LIGOScientific:2020tif,LIGOScientific:2021sio} only compute bounds on individual deviation coefficients. Better constraints from multi-parameter tests can be achieved by reparametrizing deviations in terms of uncorrelated quantities~\cite{Datta:2022izc}. Multi-parameter tests might also become effective in the future with multiband observations, where a GW signal is tracked in its evolution from the frequency band of space-based detectors to that of third-generation ground-based detectors~\cite{Datta:2020vcj}, thus covering several weeks or months of inspiral signal over a broad frequency range.

Current detectors are sensitive to signals emitted not only during the inspiral, but also during the merger and post-merger regimes. Therefore, parametric deviations in the spirit of Eq.~\eqref{eq:ppn2} have been also applied to NR-calibrated IMR waveform models, in particular to phenomenological frequency-domain waveforms. In this case, fractional deviations are instead added to the coefficients of a phenomenological phase model, tuned to NR simulations, as the merger and post-merger regimes are dominated by highly relativistic and strong-field effects not adequately captured by PN expansions. In the analyses performed by the LVK Collaboration, the deviations were added to the phenomenological coefficients of the gravitational waveform model \textsc{IMRPhenomPv2}~\cite{Hannam:2013oca,Husa:2015iqa,Khan:2015jqa}, whose merger--ringdown phase is characterised by the sets of coefficients $\beta_{i}$ and $\alpha_{i}$ (see~\cite{Khan:2015jqa} for a definition of these parameters). 

\begin{figure}[h]
\begin{center}
\includegraphics[width=1.0\textwidth]{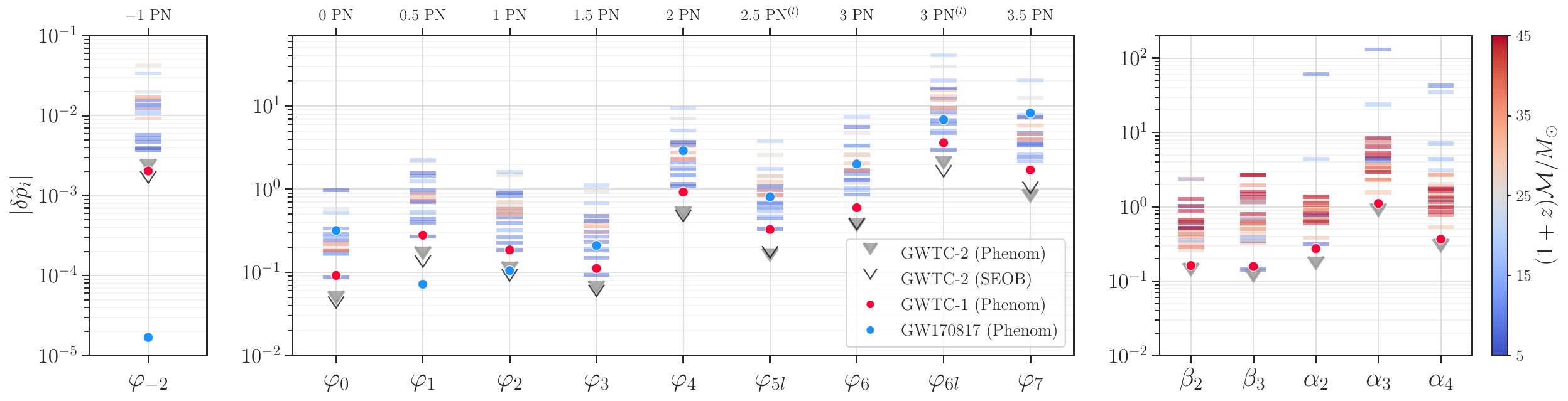}
\caption{The left two columns of the figure show 90\% upper bounds on the fractional deviation parameters $\delta\hat{\phi}_{n}$ that appear in the inspiral part of the signal. In contrast, the right column represents the constraints on deviations in the post-inspiral coefficients $\delta\beta_i$, $\delta\alpha_i$ (which should also reduce to 0 in pure GR). The horizontal stripes denote constraints from individual events calculated employing phenomenological waveform models, with their various colours indicating the corresponding chirp mass of the binary (redshifted, in the detector frame), as specified by the colour bar. Filled and unfilled triangles represent combined results from all events obtained with two approaches, one based on the phenomenological model and the other on an effective one-body model. The previous catalogue's combined results~\cite{LIGOScientific:2019} are shown as red circles for comparison. Additionally, the constraints arising from the binary neutron star event GW170817 (see Ref.~\cite{LIGOScientific:2018dkp}) are shown as blue circles. This figure is taken from~\cite{LIGOScientific:2020tif}.
\label{fig:PPN_o3a}}
\end{center}
\end{figure}

Figure~\ref{fig:PPN_o3a}, from Ref.~\cite{LIGOScientific:2020tif} shows the constraints on the deviation coefficients obtained from the GW data collected during the first, second and first half of the third observing runs of the LIGO-Virgo network (GWTC-2). Starting from left, the 90\%-confidence bounds on the fractional single-parameter deviations of the inspiral coefficients $\{\delta\hat{\phi}_{n}\}$, and post-inspiral coefficients $\{\delta\beta_{i}\}$ and $\{\delta\alpha_{i}\}$ are shown as indicated on the horizontal axis. The inspiral coefficients are constrained at each PN order starting from the ($-1$)PN order until $3.5$PN order, with deviations of the logarithmic terms in Eq.~\eqref{eq:phase_gr}, $\varphi_{5\ell}$ and $\varphi_{6\ell}$, being treated separately. Each bar corresponds to an individual event, and events are colour-coded based on their chirp mass, as indicated in the colour bar on the right. As expected, more massive events contribute predominantly to the bounds placed on the merger and post-merger deviations, since their inspiral regimes occur around the low-frequency end of the sensitivity range of current detectors. The bounds obtained with the observation of the binary neutron star event GW170817 are shown as blue circles: in this case, thousands of GW cycles fell in the frequency band of the LIGO-Virgo network, yielding particularly stringent constraints on the emission of dipolar radiation. This readily follows from  the phasing formula of Eq.~\eqref{eq:phase_gr}, where low-order PN terms dominate at low frequencies. The presence of dipolar radiation can also be constrained introducing a parametrized total emitted GW flux $\mathcal{F}=\mathcal{F}_\mathrm{GR}(1+B\, c^2/v^2)$, where $B$ represents a dipolar correction. The constraint on the parameter $B$ obtained with GW170817 is five orders of magnitude weaker than that proceeding from the double pulsar PSR J0737–3039A/B~\cite{Kramer:2021jcw} (as indirect observations of the emitted gravitational radiation from the binary's orbital period decay), due to the much longer observation period of the latter. Double pulsar and LVK constraints are complementary though, as they probe different regimes (mildly and strongly relativistic, respectively) and different sources (as LVK observations encompass also neutron-star-black-hole and binary BH systems). 

Multiple events can be statistically combined to obtain more stringent bounds on each deviation coefficient. In Fig.~\ref{fig:PPN_o3a}, combined bounds are denoted by a GWTC-* label, where the star refers to the reference catalog of transients for each bound. Combined results (using phenomenological models) from the first two observing runs (O1-O2)~\cite{LIGOScientific:2019} are shown as red circles for comparison. To gauge the impact of waveform modelling systematics, inspiral parametrized tests were performed with two different waveform models incorporating PN information. The open and closed triangles in Fig.~\ref{fig:PPN_o3a} show the combined results for all events in the figure, obtained by employing the phenomenological GW model \textsc{IMRPhenomPv2} and the effective-one-body GW model \textsc{SEOBNRv4\_ROM}~\cite{Bohe:2016gbl}, respectively.  

\subsection{Tests of the black hole nature using spin-induced quadrupole moment measurements}
\label{subsec:siqm}

Spin-induced multipole moments arise due to the spinning motion of compact objects and leave distinctive imprints on the emitted GW signal. According to the BH no-hair theorem, the multipolar structure of the Kerr metric is entirely determined by the BH's mass $m$ and adimensional spin $\chi$, with the $l$-th mass and current multipole moments, $M^{\mathrm{BH}}_{l}$ and $S^{\mathrm{BH}}_{l}$, obeying the relation~\cite{Carter:1971zc,Hansen:1974zz}:
\begin{equation}
    M^{\mathrm{BH}}_{l}+i\mkern1mu S^{\mathrm{BH}}_{l}=m^{l+1}(i\mkern1mu\chi)^{l} \; ,
\end{equation} 
where $i\mkern1mu$ denotes the imaginary unit.
From the above equation, one can deduce that the mass quadrupole moment of a rotating BH is given by $M_{2}^{\mathrm{BH}}=-m^{3}\chi^{2}$~\cite{Poisson:1997ha}. This simple relation does not apply to bodies with an internal structure such as neutron stars or more exotic objects like boson stars~\cite{Ryan:1996nk}. For generic compact objects, we may then write a generalised form for the mass quadrupole moment:
\begin{equation}
    Q\equiv M_{2}=-\kappa\,m^3 \, \chi^2\, ,
    \label{eq:siqm}
\end{equation}
where we call $\kappa$ the spin-induced quadrupole moment (SIQM) coefficient. For BHs,
$\kappa_\mathrm{BH}=1$, whereas for other compact objects like neutron stars or exotic stars this coefficient varies depending upon the internal composition or the equation of state of the object. For neutron stars, $\kappa_\mathrm{NS}\sim2-14$, whereas $\kappa$ is typically larger for boson stars~\cite{Pappas:2012qg,Pappas:2012ns,Ryan:1996nk,Vaglio:2023lrd}.

Among spin-induced multipole moments, the spin-induced quadrupole is the leading contribution in the post-Newtonian GW phasing; it appears at $2$PN order, with next-to-(next-to-)leading order effects appearing at $3$PN (resp. $3.5$PN) order. 

In light of Eq.~\eqref{eq:siqm}, Ref.~\cite{Krishnendu:2017shb} proposed a parametrized test where deviations are applied to the SIQM coefficients, $\kappa=1+\delta\kappa$ of a binary's components. Unlike the generic modifications of Subsec.~\ref{subsec:generic_modifications}, this parametrization was specifically designed to test the nature of compact objects through their deformation under the effect of rotation. For non-BH objects, the $\delta\kappa$ term will produce an extra quadratic-in-spin contribution in the phasing formula~\eqref{eq:phase_gr}, entering the coefficients $\phi_{n}$ with $n\in\{4,6,7\}$. By construction then, this parametrization only probes some specific variations of the $\phi_{n}$, not generic deviations as in the previous subsection. It is active only at specific PN orders, and it does not affect all the terms appearing at a given order (e.g., non-spinning or spin-orbit terms in the phasing formula are unchanged).

For GR BHs, $\delta\kappa=0$, and any non-zero value can hint at a deviation from the Kerr nature, requiring further focused studies to identify the exact nature of the object. The SIQM--based tests were employed to analyse GW events detected through the first three observing runs of the LVK ground-based detectors~\cite{LIGOScientific:2021sio}.

Generically, both spin-induced quadrupole moment parameters of a binary's components, $\delta\kappa_1$ and $\delta\kappa_2$, will appear in the phasing along with their mass and spin parameters, as schematically shown in Eq.~\eqref{eq:siqm}. However, at current sensitivities the simultaneous estimation of both parameters often ends up being uninformative. It was found that observational constraints on the BH nature of these objects are best placed on the symmetric combination $\delta\kappa_s=(\delta\kappa_1+\delta\kappa_2)/2$, rather than on the individual deviation coefficients $\delta\kappa_i$, while keeping the anti-symmetric combination $\delta\kappa_a=(\delta\kappa_1+\delta\kappa_2)/2$ to zero~\cite{Krishnendu:2017shb,Krishnendu:2019tjp}. 

\begin{figure}[h]
\begin{center}
\includegraphics[width=0.7\textwidth]{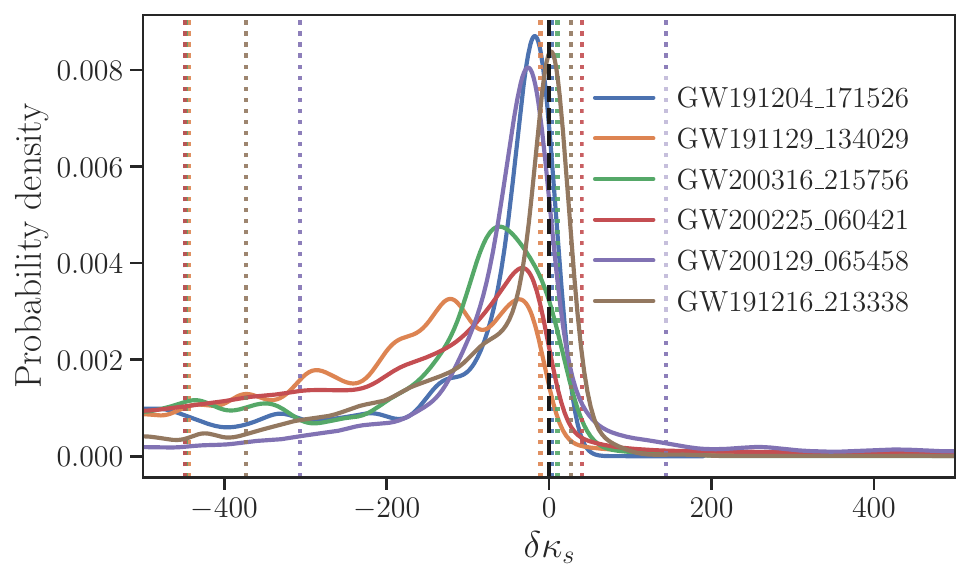}
\caption{The posteriors on the symmetric spin-induced quadrupole moment deviation parameter of a binary system $\delta\kappa_s$ obtained by analyzing a selection of the loudest events detected through the second-half of the third observing run of the LVK detectors. The 90\% confidence bounds obtained for each GW event are marked by dotted colored vertical lines. The observations are consistent with binary BHs in GR, $\delta\kappa_s^\mathrm{GR} = 0$, corresponding to the vertical black dashed line. Figure taken from~\cite{LIGOScientific:2021sio}. \label{fig:sim}}
\end{center}
\end{figure}

The spin-induced quadrupole--based test is more effective for binaries with large spin, as in this case spin-induced effects on the phasing are more significant. Fig.~\ref{fig:sim} shows the constraints on $\delta\kappa_s$ obtained with a selection of the loudest events detected through the second-half of the third observing run of the LVK detectors. 

As discussed in Ref.~\cite{Krishnendu:2019tjp}, there is a correlation between $\delta\kappa_s$ and the effective spin parameter $\chi_{\mathrm{eff}}=(m_1\chi_1^{z}+m_2\chi_2^{z})/(m_1+m_2)$, where $m_{1,2}$ are the individual masses and $\chi_{1,2}^{z}$ are the spins' components aligned with the binary's orbital angular momentum. 
This results in positive $\delta\kappa_s$ being better constrained for positive values of $\chi_{\mathrm{eff}}$, and \textit{vice versa}. This trend is evident in Fig.~\ref{fig:sim}, as the observed GW signals employed in this test mostly have small positive $\chi_{\mathrm{eff}}$ (systems with very small $\chi_{\mathrm{eff}}$ are excluded). The improved sensitivity of future interferometers and the detection of more sources with a broader range of effective spins will yield improved constraints on $\delta\kappa_s$. 

A complementary test of the nature of compact objects exploits the concept of tidal heating. This term refers to the fact that the absorption of energy and angular momentum can change the BH's mass and spin~\cite{Poisson:1994yf}. 
While the event horizon of a BH acts as a one-way surface, GWs can escape from horizonless objects, and therefore dissipation is expected to be smaller for exotic compact objects than for BHs. Different absorption properties will change the GW flux emitted by the binary, leaving an imprint on the signal we detect. Hence, constraints on the amount of tidal heating can be taken as a measure of the nature of a compact object. While tidal heating is unlikely to be measurable with current detectors, more precise measurements might be achieved with third-generation detectors~\cite{Mukherjee:2022wws}; particularly promising systems are extreme-mass-ratio-inspirals that will be observed by LISA~\cite{Cardenas-Avendano:2024mqp,Maggio:2021uge}.

\section{Polarization tests}
\label{sec:polarization}

\begin{figure}[h]
\begin{center}
\includegraphics[width=0.6\textwidth]{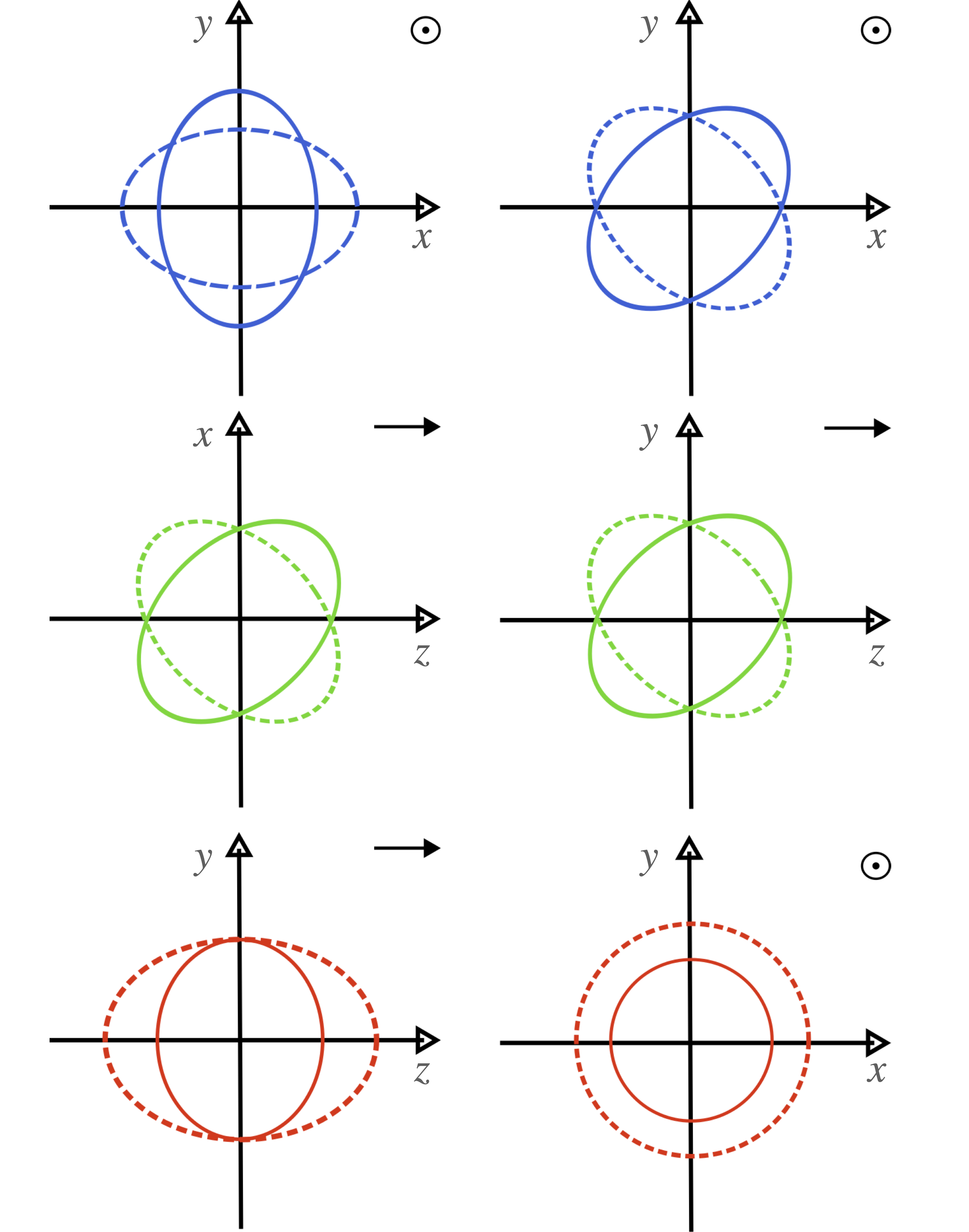}
\caption{The six GW polarizations allowed by alternative metric theories of gravity. The axes are labelled assuming that gravitational waves propagate in the $z$ direction. Each row illustrates the effect of different polarization modes on a ring of test particles: the upper row corresponds to tensor modes, the middle one to vector modes and the bottom one to scalar modes. A symbol in the upper-right corner of each set of axes indicates whether the mode is transverse (circle) or longitudinal (arrow). Adapted from~\cite{Will:2014kxa}. \label{fig:polarizations_rings}}
\end{center}
\end{figure}

Metric theories of gravity allow up to six polarization states for GWs, which can be related to the six independent components of the ``electric'' part of the Riemann tensor $R_{0i0j}$ (with $i,j$ being spatial indices)~\cite{Will:2014kxa}. These states comprise two tensorial polarizations ($h_{+}$, $h_{\times}$), two vectorial (or $x$ and $y$ modes) and two scalar modes including a transverse (also called ``breathing'' or conformal) and a longitudinal mode. Their effect on a ring of freely falling test particles is illustrated in Fig.~\ref{fig:polarizations_rings}. 

Tensor polarizations are purely transverse to the direction of propagation of the wave, whereas vector polarizations are longitudinal. GR only predicts two tensorial polarizations, corresponding to the degrees of freedom of a spin-2 massless graviton. Alternative theories of gravity will generally allow more polarization states than GR: thus, the detection of non-tensorial polarizations would provide evidence of beyond GR physics.

GW detectors at different locations have a different response to incoming signals, which will depend on their polarization and sky location. The longitudinal and breathing polarizations are not linearly independent for interferometric GW detectors, and therefore these instruments are sensitive to at most five independent polarization states~\cite{Chatziioannou:2012rf}. Assuming that a single detector output $d(t)$ is a sum of a signal $h(t)$ and a noise component $n(t)$, one can schematically write
\begin{eqnarray}
    d(t)=n(t)+F_{a}(\hat{\Omega}_s,\psi)h^{a}(t),
\end{eqnarray}
where $a$ is an index going over all the possible (non-degenerate) polarization states of $h(t)$ and $F_{a}$ are the antenna pattern functions for each polarization, which depend on the source sky location $\hat{\Omega}_s$ and polarization angle $\psi$. Generically, antenna pattern functions will be time-dependent; however, this dependence can be neglected for short-duration transients like the ones observed by current detectors. For D detectors, the equation above can be generalised to
\begin{eqnarray}
    d^{i}(t)=n^{i}(t)+F_{a}^{i}(\hat{\Omega}_s,\psi)h^{a}(t),
    \label{eq:pol_matrix_form}
\end{eqnarray}
with $i=\{1,..,D\}$. 

Several methods have been proposed to test the polarization content of GW signals. One of them relies on the concept of null streams, which will briefly outline below. The construction was initially proposed to distinguish GW bursts from transient instrumental artefacts \cite{Sutton:2009gi}. 
Eq.~\eqref{eq:pol_matrix_form} can be interpreted as a projection of the signal reaching each detector onto the hyperplane spanned by the vectors $F_{a}^{i}$. Given a certain hypothesis about the polarization content of an incoming GW signal, one can always construct an operator projecting the signal orthogonally to this hyperplane, yielding a null stream, i.e., a combination of detector outputs that does not contain any power coming from a signal with the assumed polarization content and sky location. Given a network of D detectors and $n_{\rm{pol}}$ polarizations, it is possible to construct at most D-$n_{\rm{pol}}$ null streams\footnote{%
While this statement is generally true, Ref.~\cite{Hagihara:2018azu} found that specific sky locations lead to certain polarization modes to vanish identically for all null streams, thus reducing the number of detectors needed to test independent polarization modes.%
}. 
The constraining power of different detectors is related to their relative position: the two LIGO detectors, for instance, have similar antenna pattern functions, which will result in somewhat degenerate constraints. Hence, observations involving a larger network of detectors including Virgo, KAGRA and the planned LIGO-India are particularly promising in this respect. If the excess power left after the null projection is consistent with noise fluctuations, then the data are consistent with the polarization hypothesis assumed when building the null projector. If the detector noise is taken to be stationary and Gaussian, one can then define a likelihood function for the hypothesis that the signal proceeds from a given sky location, with a given polarization content. This in turns allows to compute Bayes factors comparing different polarization hypotheses. Given a set of data streams, it is possible to construct different null projectors testing either pure~\cite{LIGOScientific:2018dkp,Pang:2020pfz} (i.e., purely scalar, vectorial or tensorial) or mixed polarizations \cite{LIGOScientific:2021sio}. 
 
Multiple GW observations can be pooled to compute combined Bayes factor for different polarization hypotheses against the purely tensorial one. Several analyses performed by the LVK Collaboration~\cite{LIGOScientific:2016lio,LIGOScientific:2018dkp,LIGOScientific:2019,LIGOScientific:2020tif,LIGOScientific:2021sio} concluded that existing observations are consistent with the signals having purely tensorial polarizations, though current conclusions are limited by the number of detectors and the SNR achievable at current sensitivities. 

The null-stream based method outlined above does not coherently track the signal's phase. An alternative approach consists in carrying out full Bayesian inference assuming a GR model for the signal $h$ and replacing the tensor antenna pattern functions with those specific to an alternative polarization content~\cite{Isi:2017fbj}. This method was employed by the LVK Collaboration to test pure polarization hypotheses for the binary BH event GW170814~\cite{LIGOScientific:2017ycc} and the binary neutron star candidate GW170817~\cite{LIGOScientific:2018dkp}. 
The latter yielded the strongest single-event constraint on the presence of non-tensorial polarizations, thanks to the precise sky localisation provided by its electromagnetic counterpart. The template-based method adopted by the LIGO-Virgo analysis yielded $\log_{10}$ Bayes factors of $20.81\pm0.08$ and $23.09\pm0.08$ in favour of tensorial polarizations over purely vectorial and scalar modes, respectively~\cite{LIGOScientific:2018dkp}. These constraints are several orders of magnitude more stringent than recent combined results using binary BH events\footnote{%
Stronger constraints on the polarization content of GW170817 were obtained restricting the inclination prior using information extracted from hydrodynamical simulations \cite{Mooley:2022uqa,Takeda:2020tjj}.%
}.  

To avoid the use of GR templates, which might not adequately represent the signal in alternative theories of gravity, a model-agnostic method representing polarization states with sums of sine-Gaussian wavelets has also been proposed~\cite{Chatziioannou:2021mi} and employed to place upper limits on the SNR of potential non-tensorial polarizations in the detector's response.

\section{Propagation tests}
\label{sec:propagation}

Like electromagnetic waves, GWs in GR propagate non-dispersively in vacuum, at the speed of light $c$. The situation might change in alternative theories of gravity, where GWs might propagate at a speed that is different from $c$ or that is a function of the wave's frequency. Different dispersion relations for GWs in theories beyond GR can be a consequence of some symmetry-breaking mechanism (e.g., due to violation of local Lorentz invariance in the new theory), additional fields or different assumptions regarding the propagation of gravitational fields. The last category includes, for instance, theories postulating the existence of extra dimensions or massive gravity~\cite{deRham:2014zqa}, which endows the graviton, i.e. the particle mediating gravitational interactions, with a non-zero mass. 

To ensure consistency with existing precision tests of gravity, such as Solar system or binary-pulsar measurements, a screening mechanism is occasionally required in order to suppress deviations from GR within a certain distance (or screening radius) from the source. In these cases, imprints of new physics might still be observed through the propagation of gravitational waves over galactic and cosmological distances. 

Though alternative theories of gravity will also predict modifications in the generation of gravitational waves, the latter can be neglected assuming that corrections to the orbital motion and radiation reaction are suppressed by some power of $\lambda/\lambda_{\rm{mod}}\ll 1$, where $\lambda$ is the gravitational-wave wavelength and $\lambda_{\rm{mod}}$ is the characteristic length associated with beyond GR effects. For instance, in massive gravity $\lambda_{\rm{mod}}$ could be identified with the graviton's Compton wavelength $h/(m_{g}c)$. On the other hand, modifications to the propagation mechanism of gravitational waves will be amplified by the distance travelled by the signal. 

Starting from this premise, Ref.~\cite{Mirshekari:2011yq} studied the propagation of gravitational waves obeying a modified dispersion relationship
\begin{eqnarray}
E^{2}=p^{2}c^{2}+A_{\alpha}p^{\alpha}c^{\alpha},
\label{eq:mdr}
\end{eqnarray}
in a Friedman-Robertson-Walker background, where $c$ is the speed of light, $p,E$ are the four-momentum and energy of gravitons and $A_{\alpha},\alpha$ are some phenomenological parameters.
The GR limit corresponds to setting $A_{\alpha}=0$, i.e. to assuming that gravitational waves travel along null geodesics. Setting $\alpha=0$ would correspond to endowing the graviton with a mass $m_g=\sqrt{A_0}/c^2$~\cite{Will:1997bb}, whereas terms with $\alpha\neq0$ model generic violations of local Lorentz invariance. 

For a generic modified theory predicting both a massive graviton and a Lorentz-violating term proportional to $A_\alpha$ with $\alpha\neq0$, gravitational waves might propagate at subluminal or superluminal speed, according to the relative magnitude of the graviton's mass and of the $A_{\alpha}$ coefficients. This can be seen by defining the graviton's (particle) velocity as $v=c^2 p/E$ and expanding Eq.~\eqref{eq:mdr} for small magnitudes of $A_{\alpha}$, yielding $v^2/c^2=1-\frac{m_g^2c^4}{E^2}-A_\alpha E^{\alpha-2}$. 

While local Lorentz invariance is one of the key ingredients of  Einstein's equivalence principle, it is not necessarily preserved in alternative theories of gravity. Theories introducing a preferred frame or direction, as it is the case for Einstein-Aether theory~\cite{Jacobson:2004ts} or Ho\v{r}ava gravity~\cite{Horava:2009uw}, are Lorentz-violating, though Lorentz-invariance can be still recovered in the low-energy limit.

The inspiral phase shift induced by the modified dispersion relation of Eq.~\eqref{eq:mdr} can be analytically computed as a function of $\alpha,A_\alpha$. We will report below the results obtained applying the SPA and assuming that GWs propagate at the particle velocity $v=c^2 p/E$~\cite{Mirshekari:2011yq}, as this is the method followed by the LVK Collaboration\footnote{%
A similar analysis can be carried out assuming that gravitational waves propagate at the group velocity, without relying on the SPA, but on the Wentzel–Kramers–Brillouin (WKB) approximation~\cite{Ezquiaga:2022nak}.%
}. Let $\phi(f)$ be phase of a gravitational wave propagating in GR, as a function of frequency. Then, 
Eq.~\eqref{eq:mdr} induces a phase shift $\phi(f)\rightarrow\phi(f)+\delta\phi_\alpha(f)$: 
\begin{eqnarray}
\delta\phi_{\alpha}(f)\propto 
\begin{cases}
\begin{array}{ll}%
\frac{D_\alpha \lambda_\mathrm{A}^{\alpha-2}}{(\alpha-1)(1+z)^{1-\alpha}}\left(\frac{f}{c}\right)^{\alpha-1},  &  \alpha\neq1 \\[10pt]
\frac{D_{1}}{\lambda_{A}}\ln{\left(\frac{\pi G\mathcal{M}f}{c^{3}}\right)},  & \alpha=1 \\
\end{array}
\end{cases}
\label{eq:mdr_shift}
\end{eqnarray}
where $\mathcal{M}$ is the detector-frame chirp mass of the binary, $\lambda_{A}\coloneqq hc\left(A_{\alpha}\right)^{1/(\alpha-2)}$ and $D_{\alpha}$ is an effective distance parameter, depending on the source redshift $z$ and on the assumed cosmological model
\begin{eqnarray}
D_\alpha\coloneqq\frac{(1+z)^{1-\alpha}}{H_0}\int_{0}^{z}{\frac{(1+z')^{\alpha-2}}{\sqrt{\Omega_{\rm m}(1+z')^3+\Omega_\Lambda}}c\,dz'},
\end{eqnarray}
where $H_0$ is the Hubble constant, while $\Omega_{\rm m}$ $(\Omega_{\Lambda})$ are the matter (dark energy) density parameters. 
Eq.~\eqref{eq:mdr_shift} shows that the dephasing acquired by gravitational waves because of the modified dispersion relation~\eqref{eq:mdr} accumulates with the distance travelled from the source to the detector. The case $\alpha=2$ yields a dephasing linear in frequency, which can always be reabsorbed in a redefinition of the coalescence time. 

Since the coalescence time is, in general, unknown, $A_{2}$ cannot be constrained in the absence of further information. An important exception is represented by sources associated to an electromagnetic counterpart, as in this case a constraint on $A_2$ can be placed by comparing the propagation speed of light with that of gravitational waves. This test was carried out for the gravitational-wave event GW170817 and associated gamma-ray burst GRB170817A~\cite{LIGOScientific:2017zic}. 

\begin{figure}[h]
\begin{center}
\includegraphics[width=0.8\textwidth]{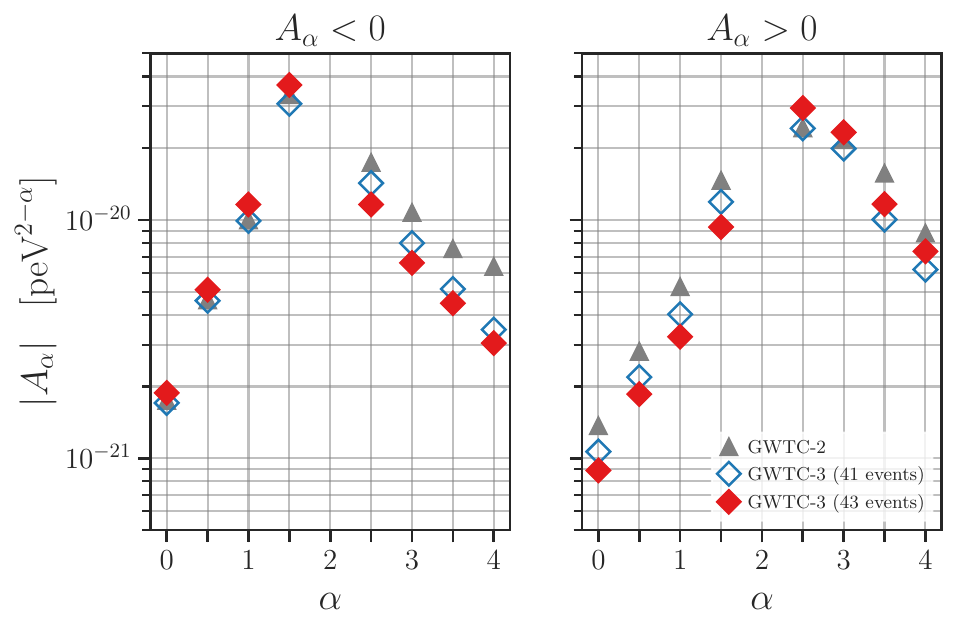}
\caption{Upper bounds on the absolute value of the $A_\alpha$ coefficients parametrizing the modified dispersion relation of Eq.~\eqref{eq:mdr}. Cases where $A_\alpha$ is positive or negative are considered separately. Figure from~\cite{LIGOScientific:2021sio}.\label{fig:liv_bounds}}
\end{center}
\end{figure}

The magnitude of the parameters $A_\alpha$ can be bounded by combining multiple gravitational-wave observations. The tests performed by the LVK Collaboration consider values of the $\alpha$ ranging from 0 to 4 in steps of $0.5$, excluding $\alpha=2$ for the reasons above. Each value is considered individually, i.e. the analysis does not take into account modified dispersion relations with both a massive graviton and other generic Lorentz-violating terms at once. The latest bounds on the coefficients $A_{\alpha}$ obtained by the collaboration \cite{LIGOScientific:2021sio} combined gravitational-wave observations from the first three observing runs and are shown in Fig.~\ref{fig:liv_bounds}. The analysis placed a bound on the graviton's mass more stringent than that obtained with Solar System measurements \cite{Bernus:2020szc}. 

The method presented above only consider isotropic and non-birefringent effects. In the most general case though, Lorentz violations could lead to anisotropy, dispersion, and birefringence in the propagation of gravitational waves~\cite{Kostelecky:2016kfm,Mewes:2019dhj}. 
Anisotropy effects are caused by terms breaking rotational invariance. Birefringence arises when the left-hand and right-hand circular polarizations of GWs, defined as the complex combinations $h_{L/R}=\frac{(h_{+}\pm i\mkern1mu h_{\times})}{\sqrt{2}}$, satisfy different propagation equations i.e., they might be travelling at different velocities (velocity birefringence) or their relative amplitude could vary while they propagate (amplitude birefringence).

Birefringent propagation has been put forward as a possible signature of CPT (Charge-Parity-Time) violation, which is closely related to local Lorentz invariance~\cite{Greenberg:2002uu}. While GR is parity-invariant, some of the theories that have been proposed to explain the nature of dark matter and dark energy, such as Chern–Simons gravity~\cite{Jackiw:2003pm}, ghost-free scalar-tensor gravity or Ho\v{r}ava-Lifshitz gravity contain parity violating terms. Searches for birefringence on open GW data have been carried out by several authors, finding no evidence of deviations from GR (e.g.,~\cite{Ng:2023jjt,Okounkova:2021xjv,Shao:2020shv,Wang:2021gqm,Zhu:2023wci}).

The amplitude of GWs in alternative theories of gravity might be also damped with respect to the GR case due to additional cosmological friction. For instance, certain classes of scalar-tensor theories can modify the propagation of gravitational waves in vacuum, either through an altered propagation speed or through an additional damping of the wave amplitude. 
This effect arises due to the interaction between the additional degrees of freedom and the curvature term~\cite{Saltas:2014dha, Lombriser:2015sxa, Dalang:2019rke, Romano:2023ozy}, and it introduces an extra damping effect in the cosmological friction term, owing to which the luminosity distance inferred from the GW observations differs from that deduced from electromagnetic observations. The ratio of the two luminosity distances is, in general, a function of the redshift, and the evolution can be modelled using two parameters, as discussed in~\cite{Belgacem:2018lbp}. In the GR limit, this ratio reduces to unity at any given redshift. Hence, such an effect can be used as a generic test for deviations from GR~\cite{Afroz:2023ndy}. This test has been proposed in the context of both current (LIGO, Virgo, KAGRA \cite{Ezquiaga:2021ayr}) and forthcoming ground-based (e.g. Einstein Telescope~\cite{Belgacem:2017ihm,Matos:2021qne,Maggiore:2019uih}) and space-based (e.g. LISA~\cite{LISACosmologyWorkingGroup:2022jok,LISACosmologyWorkingGroup:2019mwx,LISA:2022kgy}) GW detectors.

Other scenarios where the two luminosity distances can differ are those where gravitons can decay into lighter particles (e.g. photons \cite{deRham:2016nuf}) or leak into extra spatial dimensions. These mechanisms would alter the scaling of the GW strain amplitude $h$ with the luminosity distance $d_{\mathrm{L}}$. In GR, one has $h\propto1/d_{\mathrm{L}}$; the scaling would be steeper in the presence of extra dimensions, leading GR templates to systematically overestimate the distance of sources~\cite{Deffayet_2007}. If the presence of extra dimensions does not affect the propagation of light, the electromagnetic and GW luminosity distance will differ. These two quantities can be independently estimated when a GW event is accompanied by an electromagnetic counterpart, as in the case of GW170817~\cite{LIGOScientific:2017zic}. 

In the absence of an electromagnetic counterpart, the redshift of the sources might still be estimated at a population level using techniques such as the statistical host identification method \cite{Schutz,Gray:2019ksv,MacLeod:2007jd}, measuring cross-correlations with galaxy surveys~\cite{Oguri:2016dgk,Bera:2020jhx,Mukherjee:2020hyn} or using source-frame mass models to break the mass-redshift degeneracy in the detector frame~\cite{Mastrogiovanni:2021wsd,Ezquiaga:2022zkx}. These methods allow to perform tests of GW propagation with so-called dark sirens, i.e., GW events without any detected electromagnetic counterparts \cite{Mukherjee:2020mha, Mancarella:2021ecn, MaganaHernandez:2021zyc}.  

Other approaches involving dark sirens to test alternative gravity theories at cosmological scales include the study of lensed gravitational-wave signals produced by such sources while passing through the intermediate clustered matter \cite{Mukherjee:2019wcg, Balaudo:2022znx, Ezquiaga:2020dao}.

\begin{figure}[h]
\begin{center}
\includegraphics[width=0.8\textwidth]{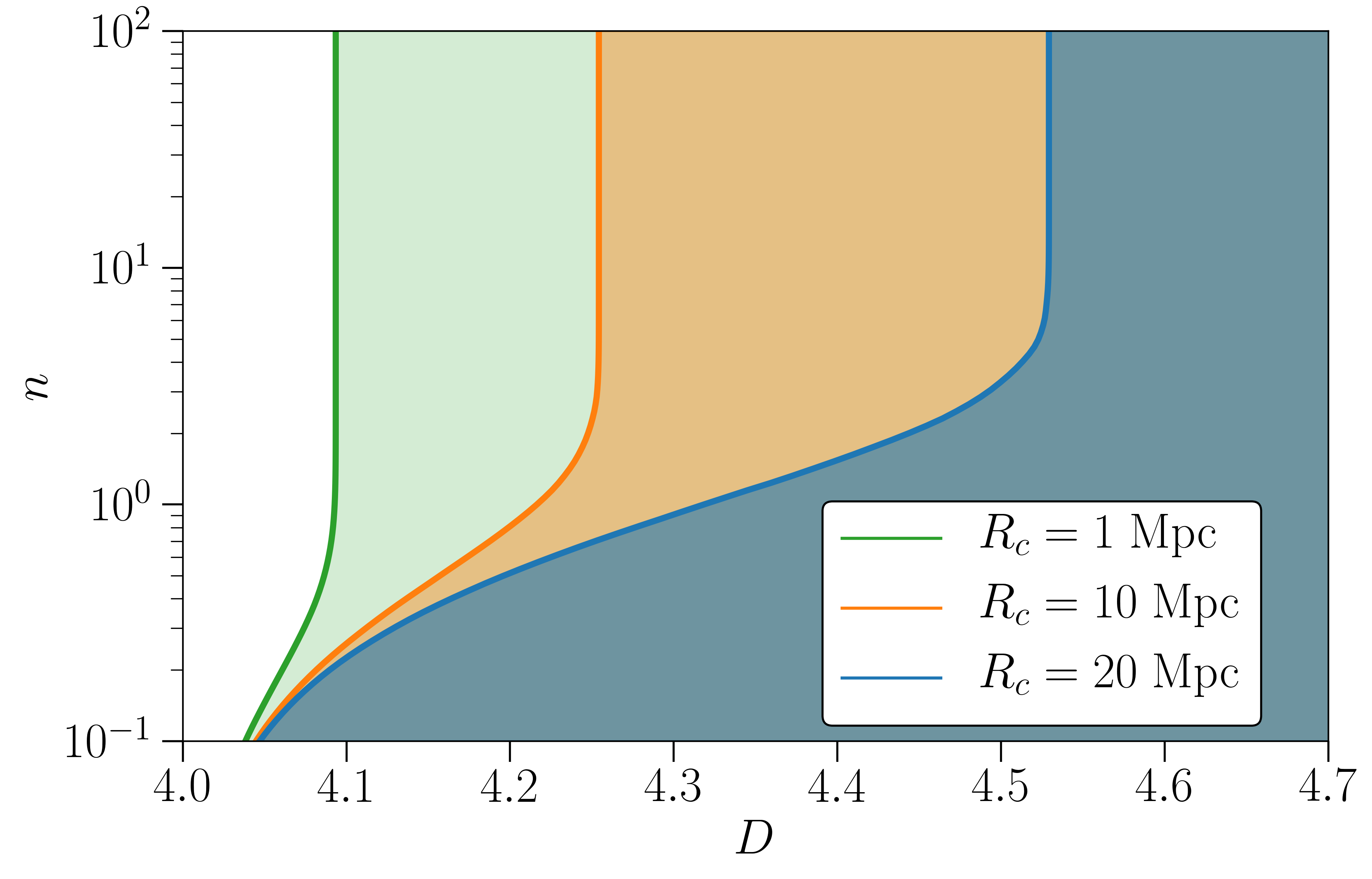}
\caption{Constraints on the number of extra spacetime dimensions obtained with GW170817. 90\% upper bounds on the number of spacetime dimensions $D$, for different values of the screening radius $R_{\rm{c}}$ and transition steepness $n$, as per Eq.~\eqref{eq:liv_extraD_scaling}. Shaded areas indicate regions excluded by the data. Taken from \cite{LIGOScientific:2018dkp}.\label{fig:extraD_bounds}}
\end{center}
\end{figure}

Assuming that deviations from GR become appreciable only beyond a certain distance $R_{\rm{c}}$ from the source, Ref.~\cite{LIGOScientific:2018dkp} tested the presence of extra spatial dimensions through a modified scaling of the GW strain amplitude
\begin{equation}
    h\propto \frac{1}{d_{\rm{L}}}\left[1+\left(\frac{d_{\rm{L}}}{R_{\rm{c}}}\right)^{n}\right]^{\frac{4-D}{2n}},
    \label{eq:liv_extraD_scaling}
\end{equation}
where D denotes the number of spacetime dimensions and the integer number $n$ models the steepness of the transition between the screened and unscreened regions. In the above, $d_{\rm{L}}$ coincides with the electromagnetic luminosity distance which is assumed to be unaffected by the leakage into extra dimensions. Notice that, by construction, the above equation gives the expected scaling of $h$ when specialized to four dimensions. For a fixed transition steepness, the upper bound on the number of dimensions $D$ depends on the screening radius, with smaller screening radii yielding more stringent bounds, as shown in Fig.~\ref{fig:extraD_bounds}. The results obtained so far are consistent with the spacetime being four-dimensional~\cite{LIGOScientific:2018dkp,MaganaHernandez:2021zyc}. 

Similarly, Ref.~\cite{Pardo:2018ipy} tested the existence of decay channels for the graviton through an exponentially damped amplitude scaling, finding a lower bound on the graviton's lifetime of $4.5\times10^{8}$ yrs. 

In the presence of multiple tensor fields interacting with the massless graviton predicted by GR (for instance, in the context of bigravity or Yang-Mills theories), there will be other interesting propagation effects such as GW oscillations, due to the mixing of different metric perturbations~\cite{Max:2017flc}, GW echoes, caused by the decoherent propagation of different tensor modes~\cite{Ezquiaga:2020dao}, and amplitude and phase distortions of the GW signal~\cite{BeltranJimenez:2019xxx, Ezquiaga:2021ler}.

\section{Tests of the nature of the remnant black hole}
\label{sec:remnant}

Tests of the predictions of GR regarding the nature of BHs based on the GW radiation from binary BH mergers may also be performed by focusing on the post-merger phase of the signal. These tests have the advantage of only relying on basic properties of the remnant object, checking its Kerr BH nature expected in GR, without requiring a knowledge of the parameters of the progenitor system and without depending on a model for the entire waveform. They are, however, challenging in practice due to the sensitivity limitations of the detectors, as they only focus on the decaying late-time train of the GW emission, typically with a much reduced SNR compared to the whole inspiral-merger-ringdown signal. These tests can be divided into two main categories: the spectral analysis of the QNMs of the remnant in the ringdown, and the search for possible echoes of the GW emission during and after the ringdown due to a non-GR imperfectly absorbing remnant horizon. These remnant tests, as well as the theory and computation of QNMs, are the subject of Chapter 9, and we shall only briefly discuss them here for completeness --- referring the reader to the aforementioned chapter for a much more comprehensive treatment and review.

\subsection{Ringdown tests}
\label{subsec:QNMs}

After the two progenitor BHs merge, the emitted gravitational radiation rapidly decays as the remnant settles towards its stationary final state. At late enough times, the emitted GWs can be described as a discrete sum of exponentially decaying modes with constant-frequency oscillations: the QNMs of the final BH (see, e.g., \cite{Kokkotas:1999,Berti:2009}). These modes can be described and computed in linear perturbation theory around the final state~\cite{Teukolsky:1973,Leaver:1985}. In GR, as a consequence of the BH no-hair and uniqueness theorems, and in the absence of electric charge for astrophysical objects, this final state is a Kerr BH, entirely characterized by two parameters, its mass and its dimensionless spin magnitude. Hence, the QNM frequencies and damping times also solely depend on these two parameters. \textit{Black hole spectroscopy} aims at testing this fundamental prediction by extracting at least two different QNMs from a ringdown signal and checking the consistency of their frequencies and damping times with the GR values arising from the two parameters of a single Kerr remnant.

The dominant QNM in a typical ringdown from a binary merger is the longest-lived (fundamental) mode of the quadrupolar angular harmonic of the signal. For tests that remain independent of the knowledge of the entire waveform, another, subdominant mode must be detected as well. Different strategies --- depending on the merger's configuration~\cite{JimenezForteza:2020,Ota:2020} --- focus on attempting to measure either the fundamental mode of a higher multipole, or one of the much faster decaying component modes (\textit{overtones}) of the quadrupolar harmonic. Identifying such a subdominant QNM in current GW observations remains very challenging either way. Systematic analyses performed by the LVK Collaboration looked for evidence of subdominant QNMs in the ringdown of events from the first three observing runs~\cite{LIGOScientific:2020tif,LIGOScientific:2021sio}, only finding weak evidence of overtones in the loudest signals. Likewise, other attempts so far have been focusing on the very loudest available ringdown signals such as that of the first GW event, GW150914~\cite{LIGOScientific:2016lio,Isi:2019,Carullo:2019,CalderonBustillo:2021,Cotesta:2022,Finch:2022,Ma:2023,Isi:2023,Wang:2023,Carullo:2023}, that of GW190521~\cite{LIGOScientific:2020a,Capano:2023,Siegel:2023}, and a few others (e.g.~\cite{Gennari:2023}, with weak evidence for subdominant harmonics). One may also combine multiple loud-enough events in order to obtain joint constraints on deviations from the general-relativistic QNM spectrum~\cite{Yang:2017,LIGOScientific:2020tif,LIGOScientific:2021sio}. A strategy based on the expected relative amplitudes and phases of multiple QNMs in the ringdown resulting from a binary BH merger, rather than their frequencies and damping times, has also been proposed~\cite{JimenezForteza:2023}. Results of QNM-based tests so far have been compatible with the GR expectations, but still bear large statistical uncertainties; this should improve substantially with further increases in sensitivity in current and future detector networks~\cite{Cabero:2020,Bhagwat:2023}.

An alternative approach, intermediate between the IMR consistency tests as presented in Sec.~\ref{subsec:imr_consistency} above
(of which it may be considered a variant) and the ringdown-only approaches discussed here, is possible, avoiding the need for subdominant rindgdown modes at the expense of recovering a dependence in the modelling of the entire waveform.
It consists in extracting the remnant’s mass and spin parameters separately from only the dominant mode of the ringdown, and from the inspiral-merger part of the waveform, checking their consistency with a single Kerr final BH.
This can be performed, e.g., using parameterized IMR waveforms in the effective-one-body formalism~\cite{Ghosh:2021}. Such tests have been carried out by the LVK Collaboration for the events detected so far in the first three observing runs, showing no evidence for deviations from GR at this stage~\cite{LIGOScientific:2016lio,LIGOScientific:2020tif,LIGOScientific:2021sio}. Information from the pre-merger part of the signal may also be used to constrain the amplitudes of the ringdown modes, using the calibration of these amplitudes from numerical IMR waveforms. This leads to other pre-merger--informed tests of deviations to GR that do not require the identification of subdominant modes in the ringdown, as performed by the LVK Collaboration with the Python code \textsc{pyRing}~\cite{LIGOScientific:2020tif,LIGOScientific:2021sio}, or by~\cite{Gennari:2023} with a parametrized effective-one-body model.

\subsection{Searches for echoes}
\label{subsec:echoes}

The post-merger part of GW signals also carries information about the presence of an event horizon around the final object, as expected in GR when the progenitors and remnant are massive enough to necessarily be BHs rather than neutron stars. Various alternative, non-GR, horizonless objects have been proposed in the literature (see section 9.5.3 in Chapter 9, and references therein). The GW signature of such an object being formed as the result of a compact binary merger can be the presence of additional, very long-lived QNMs~\cite{Chandrasekhar:1991,Franzin:2024}, or that of proper \textit{echoes} --- lower-amplitudes repetitions of the final part of a typical GW signal at later times due to multiple successive reflections of the infalling gravitational radiation on the remnant object, lacking a perfectly absorbing horizon~\cite{Kokkotas:1995,Tominaga:1999}.

Searching for such echoes has become a standard test, which has been performed by the LVK Collaboration on the events detected so far in the first three observing runs. Results were negative and were used to set upper limits on GR deviations~\cite{LIGOScientific:2020tif,LIGOScientific:2021sio}, while some independent analyses have claimed evidence for echoes in some of the signals, leading to a controversy (see, e.g., \cite{Abedi:2017,Conklin:2018,Abedi:2019,Holdom:2020,Westerweck:2018,Lo:2019,Tsang:2020}, and further details and references in section 9.6.4 of Chapter 9).

\section{Challenges and future prospects}
\label{sec:challenges}

The tests performed on GW data so far have not shown any statistically significant evidence of deviations from GR. Moreover, they placed stringent constraints on viable alternative theories of gravity~\cite{Langlois:2017dyl,LIGOScientific:2018dkp}. At present, the power of these tests is somewhat limited by the modest SNR achievable at current sensitivities, resulting in relatively wide posterior distributions for the parameters to be constrained. The interpretation of the results can be also hindered by non-Gaussian and non-stationary features in the detectors' noise, which might be mistakenly taken as deviations from GR if they are not accounted for~\cite{Kwok:2021zny}. Such features are not uncommon in practice, with a prominent example being the loud glitch affecting the LIGO-Livingston detector just before the merger time of GW170817~\cite{LIGOScientific:2017zic}. GW data affected by these issues need to be treated with care. The power coming from instrumental artefacts can be subtracted from the data following a method known as \textit{deglitching}~\cite{Davis:2022ird}; an alternative route is that of performing Bayesian inference on the signal and glitch simultaneously~\cite{Ashton:2022ztk,Chatziioannou:2021ezd,Dideron:2022tap,Plunkett:2022zmx}.
  
Third-generation and space-based GW observatories offer prospects of performing tests of GR with an exquisite precision, unattainable by current ground-based instruments. The signals observed will be louder and more numerous, sharpening both individual and combined constraints that can be placed on possible deviations from GR~\cite{Maggiore:2019uih}. Detectors operating in complementary frequency bands will enable multiband observations~\cite{Sesana:2016ljz}, which might help to break degeneracies between parametric deviations from GR and the binary's intrinsic parameters~\cite{Gupta:2020lxa}, yielding improved constraints~\cite{Barausse:2016eii,Chamberlain:2017fjl}.

At the same time, the increased sensitivity of next-generation observatories will pose entirely new challenges to testing GR frameworks. The detection rate of compact-binary coalescences will grow significantly, with the possibility of having overlapping signals in the detectors' frequency band~\cite{Regimbau:2009rk}, a demanding task for current parameter estimation techniques~\cite{Samajdar:2021egv}. GW sources will be observed with much higher SNRs: this will drastically reduce statistical uncertainties, enhancing the impact of systematic errors due to waveform inaccuracies (e.g., due to the omission of environmental effects~\cite{Cardoso:2019rou,CanevaSantoro:2023aol}, or the approximated treatment of precession, eccentricity, etc.~\cite{Saini:2022igm}), but also to the finite resolution of NR waveforms employed in the calibration~\cite{Jan:2023raq}. These issues will need to be tackled to avoid claiming false deviations from GR~\cite{Hu:2022bji}.

\begin{acknowledgement}
The authors are especially grateful to Apratim Ganguly for reviewing the manuscript. The authors also thank Sajad Bhat, Gregorio Carullo, Jose Maria Ezquiaga, Vasco Gennari, David Keitel, Suvodip Mukherjee and Elise Sanger for useful comments. X. Jim\'enez Forteza is especially thankful to Shilpa Kastha for sharing the data required to visualize the BH area theorem test. M.C. acknowledges funding from the Spanish Agencia Estatal de Investigaci\'{o}n, grant IJC2019-041385. This work was supported by the Universitat de les Illes Balears (UIB); the Spanish Agencia Estatal de Investigaci\'{o}n, grants PID2022-138626NB-I00, PID2019-106416GB-I00, RED2022-134204-E, RED2022-134411-T, funded by MCIN/AEI/10.13039/501100011033; the MCIN with funding from the European Union NextGenerationEU/PRTR (PRTR-C17.I1); Comunitat Autònoma de les Illes Balears through the Direcci\'{o} General de Recerca, Innovaci\'{o} I Transformaci\'{o} Digital with funds from the Tourist Stay Tax Law (PDR2020/11 - ITS2017-006), the Conselleria d’Economia, Hisenda i Innovaci\'{o} grant number SINCO2022/6719, co-financed by the European Union and FEDER Operational Program 2021-2027 of the Balearic Islands; the “ERDF A way of making Europe.” LIGO Laboratory and Advanced LIGO are funded by the United States NSF as well as the Science and Technology Facilities Council (STFC) of the United Kingdom, the Max-Planck-Society (MPS), and the State of Niedersachsen/Germany for support of the construction of Advanced LIGO and construction and operation of the GEO600 detector. Additional support for Advanced LIGO was provided by the Australian Research Council. Virgo is funded, through the European Gravitational Observatory (EGO), by the French Centre National de Recherche Scientifique (CNRS), the Italian Istituto Nazionale di Fisica Nucleare (INFN) and the Dutch Nikhef, with contributions by institutions from Belgium, Germany, Greece, Hungary, Ireland, Japan, Monaco, Poland, Portugal, Spain. KAGRA is supported by Ministry of Education, Culture, Sports, Science and Technology (MEXT), Japan Society for the Promotion of Science (JSPS) in Japan; National Research Foundation (NRF) and Ministry of Science and ICT (MSIT) in Korea; Academia Sinica (AS) and National Science and Technology Council (NSTC) in Taiwan. This material is based upon work supported by NSF’s LIGO Laboratory which is a major facility fully funded by the National Science Foundation.
\end{acknowledgement}


\end{document}